# Social Influence-based Attentive Mavens Mining and Aggregative Representation Learning for Group Recommendation

Peipei Wang[1], Lin Li[1,*], Yi Yu[2], Guandong Xu[3]
[1]School of Computer Science and Technology, Wuhan University of Technology, Wuhan, China
[2]Department of Informatics, the Graduate University for Advanced Studies (SOKENDAI), Japan
[3]School of Computer Science and Advanced Analytics Institute at University of Technology Sydney, Australia
Email: [1]{ppwang07, cathylilin}@whut.edu.cn, [2]yiyu@nii.ac.jp, [3]Guandong.Xu@uts.edu.au

*Abstract*—Frequent group activities of human beings have become an indispensable part in their daily life. Group recommendation can recommend satisfactory activities to group members in the recommender systems, and the key issue is how to aggregate preferences in different group members. Most existing group recommendation employed the predefined static aggregation strategies to aggregate the preferences of different group members, but these static strategies cannot simulate the dynamic group decision-making. Meanwhile, most of these methods depend on intuitions or assumptions to analyze the influence of group members and lack of convincing theoretical support. We argue that the influence of group members plays a particularly important role in group decision-making and it can better assist group profile modeling and perform more accurate group recommendation. To tackle the issue of preference aggregation for group recommendation, we propose a novel attentive aggregation representation learning method based on sociological theory for group recommendation, namely SIAGR (short for "Social Influence-based Attentive Group Recommendation"), which takes attention mechanisms and the popular method (BERT) as the aggregation representation for group profile modeling. Specifically, we analyze the influence of group members based on social identity theory and two-step flow theory and exploit an attentive mavens mining method. In addition, we develop a BERT-based representation method to learn the interaction of group members. Lastly, we complete the group recommendation under the neural collaborative filtering framework and verify the effectiveness of the proposed method by experimenting.

## I. INTRODUCTION

With the rapid development of the information age, the information overload issue is becoming increasingly serious because of the explosive growth of information source. Recommender systems can mitigate effectively this issue and have been widely deployed in various domain, such as social networks platform (Meetup), e-commerce website (Amazon), news portal (Yahoo) and so on. A series of effective recommendation algorithms have been proposed, such as collaborative filtering [1, 2], matrix factorization [3,4], etc. Most of them focus on individual users and have certain limitations on group recommendation research [5]. However, as social animals, human beings have various group activities throughout their daily life, such as watching movies with colleagues, traveling with friends, dining with classmates and so on. Therefore, group recommendation has gained increasing attention over the last years and starting to be deployed within various use cases [6]. An effective group recommendation method not only facilitates group decision-making, but also improves user participation in Web services [5]. In this work, we mainly recommend items and services for groups that are of interest or need.

Owing to the group recommendation task expands individual users into groups, how to trade-off the preferences of different group members to make recommendations is a great challenge both academic and industry for the recommender systems. Existing preference aggregation strategies mainly rely on predefined strategies, such as average [7], least misery [8], maximum satisfaction [9] and so on. In essence, different group members may have their own specialties and show different influences according to their specialties. To a large extent, this is a dynamic process, so most methods cannot effectively complete the dynamic group decision-making process. In group recommendation task, group member influence analysis is the key to dynamic group decision-making. In previous studies, most of the analysis of group member influence is based on intuitions or hypothesis to make the analysis [5, 10, 11]. A common key intuition is that "group members tend to follow the opinions of the most important members (leaders/mavens) rather than consider opinions from all group members equally [12]." However, there is no specific theoretical basis to support the feasibility of intuitions. The research of group recommendation involves data mining, machine learning, sociology, psychology and other fields [13], so it is so significant to find a specific theory and support for the scientific method. Group modeling and group-item interaction modeling are two key elements to evaluate the group's preferences for items. In the process of group modeling, most of the work is represented based on general embedding method, which only represents individual user in the group, and ignores the interactive relationship among members in the group.

In this work, we mainly solve two basic problems in group recommendation task, namely group member influence analysis and group modeling. Distinct from the prior group preference aggregation based on predefined strategies, we propose a representation learning method based on dynamic aggregation strategy in group modeling. On the whole, first of all, it analysis the influence of group members in terms of social identity theory [14] and two-step flow theory [15] and these are theoretical support instead of the previous influence analysis based on intuitions or hypothesis. Then based on two important theories of sociology, we design a neural attention network to learn the weight of group members, it is different from the general way of weighting. In the process of group interaction with the items, every group members are assigned different weights, in this way, can complete dynamic mining mavens in the group members, to capture the complex process of group decision-making. More importantly, in order to solve the issue of general embedding and it lacks of interaction between the group members, we propose an attentive aggregation embedding representation learning method based on attention mechanism and bidirectional encoder representation from transformers (BERT) for group modeling. We regard the group as a sentence and the group is composed of words, then using the BERT generated sentence vector directly to modeling of group. Since BERT itself can jointly depend on the context, the interactive relationship between group members can be fully reflected, and then the mavens representation of dynamic mining and the whole group representation generated by BERT that can be aggregated embedding to complete the final group modeling. In addition, based on the above methods, we finally deploy neural collaborative filtering (NCF) framework to implement the whole process of group recommendation, and verify the effectiveness of the proposed method in different data sets.

In summary, the main contributions of this paper are three-fold:

- To best our knowledge, this is the first work that applying the sociological theory and deep learning to integrate in group recommendation.

- We propose a novel dynamic mavens mining method based on sociological theory and attention mechanism. Specifically, this method is based on social identity theory and two-step flow theory and attention mechanism to analyze the influence of group members, and mining the relevant mavens in the group decision-making process.

- We propose an attentive aggregation embedding representation learning method based on BERT for group modeling. This method can fully consider the interaction among group members, and its effectiveness is verified by experiments.

## II. RELATED WORK

In this section, we review the literature on group recommendation and deep learning techniques for recommender systems.

### A. Group Recommendation

In recent years, group recommendation has been widely concerned and deployed in various fields, such as movies [16], tourism [17], restaurants [18], music [19] and so on. Existing group recommendation methods can be categorized into memory-based and model-based methods [20]. Memory-based methods can be further divided into two categories, namely preference aggregation and score aggregation. Preference aggregation strategy is to first create group profile by combining all users' preferences, and then generate recommendation for the group [4, 21, 22], whereas the score aggregation strategy first predicts the score of each group member for the item, and then combines the individual scores of group members to generate recommendation [7, 23, 24]. Both aggregation strategies are based on predefined strategies (e.g., average, least misery, maximum satisfaction, etc.) and do not simulate dynamic interactions of preferences among group members. Taking average and least misery strategy as examples, average strategy takes the average scores of group members as the final recommendation scores, but may return items that favorable to some members and unfavorable to others, whereas misery strategy is to make each member of the group choose the lowest score in individual scores as the final score, but may eventually recommend some mediocre items is neither like nor dislike.

Distinct from memory-based methods, many model-based methods have also been proposed. In theoretical applications, Carvalho et al. [25] applied game theory to group recommendation, which regards group events as a non-cooperative game and suggest recommendation goal based on Nash equilibrium, but this strategy cannot recommend specific projects. Probability models are also widely applied to solve the problem of group recommendation, Liu et al. [26] proposed PIT model and it assumed that the members with the greatest influence in the group play an important role in group decision-making, group modeling is implemented by considering the personal preferences and influences of the group members. Besides, Yuan et al. [5] proposed COM model to simulate group decision-making process and generate recommendation results for the group. Similar to our proposed methods are AGR [10] and AGREE [11], both methods apply attention mechanism based on relevant intuitions or assumptions for group embedding representation. However, different from our work, on the one hand, we have the important theories of sociology as the support and apply the attention mechanism to analyze the influence of group members, and complete the dynamic mavens mining in the group. On the other hand, we propose a novel BERT representation learning and aggregation embedding to capture the modeling of group profile, which fully reflects the interaction between group members.

### B. Deep Learning for Recommender Systems

Recently, deep learning has made breakthroughs in computer vision and natural language processing [27,28,29], presented new opportunities to recommender systems. On the one hand, deep learning can learn the essential features of data from samples and obtain the deep representation of users and items. On the other hand, deep learning can carry out automatic feature learning from multi-source heterogeneous data [30,31], and obtain unified representation of data [32,33]. A large number of literatures [34, 35, 36] have applied deep learning technology to the recommender systems for different recommendation tasks. However, deep learning techniques are rarely used in group recommendation tasks. Thereinto, the proposed NCF by He et al. [37] has been successfully extended to attribution-based social recommendation [38] and item-based recommendation [39]. Literature [10, 11] is a typical example of group recommendation using attention mechanism. In our work, the first analyze the group members influence based on sociological theory, where more details will be described in Section III.B. At the same time, attention mechanism is applied to mining the dynamic mavens of group members, in particular, we regard group as a sentence [11], using a novel BERT to represent group. Then group representation and mavens representation are aggregated for group profile modeling, eventually, in the framework of NCF, we suggest the final recommendation goals.

## III. METHOD

Generally speaking, our proposed SIAGR method consist of three components: 1) attentive mavens mining which represents mavens of group based on attention mechanism and sociological theory. 2) BERT-based aggregation embedding representation learning for group profile modeling which considers interaction of group member and represents group preference. 3) generated recommendation list for groups in the framework of NCF. We first present the notations and then formulate the problem of group recommendation in this section. We then introduce the key ingredients of our proposed SIAGR method respectively. At last, we suggest the recommendation goals of groups in NCF.

## A. Notations and Problem Formulation

We use bold capital letters (e.g., X) and bold lowercase letters (e.g., x) to represent matrices and vectors, respectively. We use squiggle capital letters (e.g., $\mathcal{X}$) to denote sets. All vectors are in column forms if not clarified.

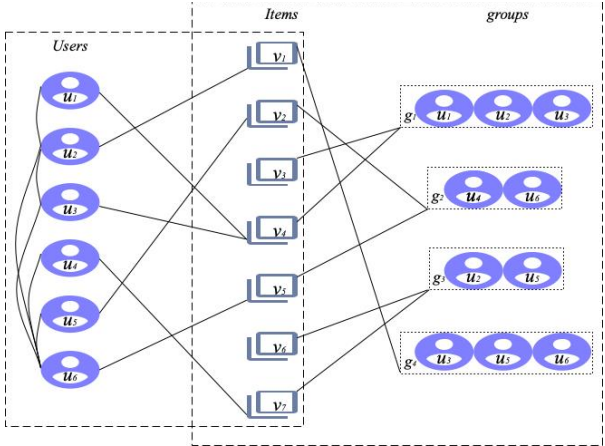

Fig. 1: Illustration of the input data for group recommendation task, including user-item interactions, group-item interactions and user-user interactions

We assume that there are a set of users $\mathcal{U}$, a set of groups $\mathcal{G}$ and a set of items $\mathcal{V}$ in group recommendation. And there are three interaction among these data, namely user-item interaction P, group-item interaction Q, user-user interaction R. We use P, Q, R represent user-item interaction, group-item interaction and user-user interaction respectively. Figure 1 illustrates the input data of our group recommendation task. Then given a target group $g_t$, our task is to generate the recommended items list for group that may be interested in, and the problem of group recommendation can be defined as follows:

Input: A set of users $\mathcal{U}$, a set of items $\mathcal{V}$, a set of group $\mathcal{G}$, user-item interactions P, group-item interactions Q, user-user interactions R.

Output: A personalized ranking function that maps an item to a ranking score for target group $f_g : \mathcal{V} \to \mathbb{R}$.

## B. Sociological Theory for Our SIAGR

Group decision-making is related closely to sociology and psychology [13]. Specially, we propose an attentive mavens mining method based on sociological theory to deal with the above problem and namely social identity theory two-step flow theory. Distinct from previous work based on intuitions [5, 10, 11], these are two important theories of sociology and can make an explanation and support from sociology for the group members influence analysis. We introduce these theories in detail how to explain and support for the group members influence below.

**Group definition:** Group is an interdisciplinary concept, many subjects define group from different perspectives. Group has drawn significant attention in the field of sociology, Ark V [40] defined that group is not the sum of individuals, group members must exist interaction and influence each other. Based on this definition of group, we can hold that group members have different influence in our work.

**Social Identity Theory:** Social identity theory is the most influential theory of group relationship [14], which was supposed by Tajfel and replenished by Turner. It shows that individuals consciously align their attitude and cognition with the group if they identify with the group, and if not identify, they break away from the group or find ways to identify positively [14]. Based on this theory, we can hold that as long as the individual identify with the group, the individual keep uniform cognition with the group.

**Two-Step Flow Theory:** Two-step flow theory shows the differences of individual influence and it was supposed by Elihu Katz [41]. This theory shows that the dissemination of infor-mation is a process from information to opinion leaders to general individuals [15]. Then, Everett Rogers [42] defined the influences, which can change other individuals' opinion in some ways. Influences are also called opinion leader, experts or mavens and we call it mavens in this work. Based on this theory, we can hold that mavens of group have a key influence on group decision-making.

Overall, through the illuminations above we can hold that individuals have different influences in the group, there are always mavens in the group and they play a key role and have a lager influence in the group decision-making. Meanwhile, group members follow the mavens' opinion to some extent.

## C. Attentive Mavens Mining and Aggregative Group Representation Learning Based on Sociological Theory

We propose an attentive aggregation representation learning method based on sociological theory for group recomme-ndation, namely SIAGR, short for Social Influence-based Attentive Group Recommendation, to address the problem of group recommendation task under representation learning framework. The process of aggregation embedding repre-sentation learning has two steps on the whole and we present the proposed method in detail as follows.

**An attentive mavens mining method based on sociological theory:** According to the analysis in section III.B., we argue that group members have different influence and mavens of group have a large influence. Take this as a theoretical support , we propose a mavens mining method based on attention mechanisms and each member has their contribution in group decision-making. In the representation learning framework, embedding $u_j$ and embedding $v_t$ are user's historical prefe-rence and target item's property respectively. $\alpha(t, j)$ represents the parameterization of neural attention network. $\mathbf{H_v}$ and $\mathbf{H_u}$ are weight matrices of the attention neural network, which convent item embedding and user embedding respectively. And $\mathbf{A}$ is a weight vector, $\mathbf{b}$ is the bias vector.

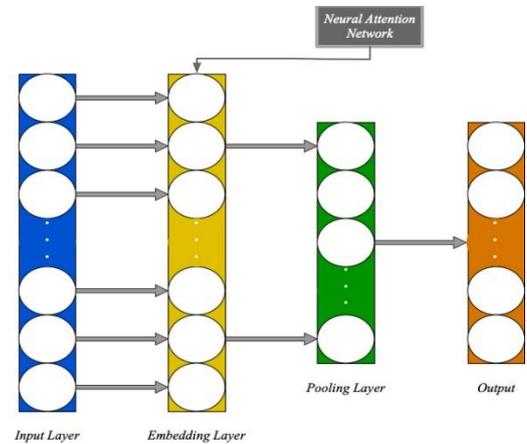

Fig. 2: Illustration of attentive mavens mining method based on sociological theory

Specifically, we use the Softmax activation function to obtain the final attention weights and it makes the neural attention network have a probabilistic interpretation. We compute the attention score as:

$$z(t, j) = A^T \text{ReLU}(H_v v_t + H_u u_j + b)$$
$$\alpha(t, j) = \text{Softmax}(z(t, j)) = \frac{\exp z(t, j)}{\sum_{j' \in o_l} \exp z(t, j')} \quad (1)$$

Figure 2 illustrates the process of attentive mavens mining in our work. With the attention mechanism, each member can contribute in group decision-making. A good example to illustrate this point as shown in Figure 3, users 1 and 12 have the largest attention weights or the largest influence as mavens in group 6, which are indicated by their darkest cells. More importantly, this approach corresponds to our theoretical support. Through our method, mavens of group can mining effectively and group members' different influence can also reflect clearly.

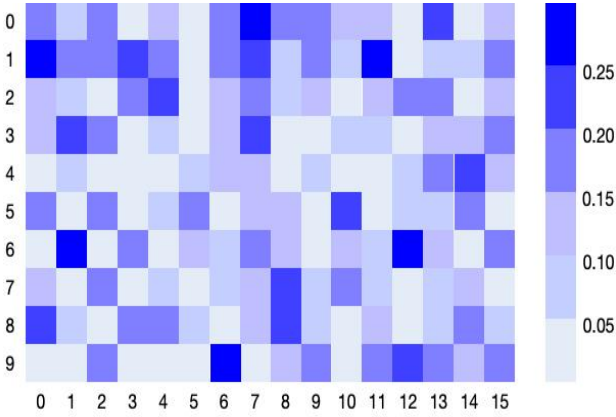

Fig. 3: Visualization for the sampled 10 groups w.r.t. attention weights, where x-axis denotes the group member-ID and the y-axis denotes the group-ID

**A BERT-based aggregation embedding representation learning method for group modeling:** This method has two components, namely BERT-based group preference embedding and aggregative representation learning with attentive mavens mining. BERT is bidirectional encoder representation from transformers [44] and it is a novel sentence-level linguistic model, we use sentence vector by BERT as whole group modeling in this work. Firstly, we can regard group as a sentence [11] and sentences are composed of words that contain the features of words. Just as the features of group members are included in the group. What is worth mentioning, A word is not a group member and may have several words represent group member. Different from linguistic model, it can consider the context well and directly obtain a unique vector representation of an entire sentence that not global pooling with weights in each layer. Hence we use BERT to obtain the sentence vector that is the group vector. By this means, it reflect the interaction between the members of the group adequately. Figure 4 illustrates the process of BERT-based attentive aggregation embedding representation learning in our work.

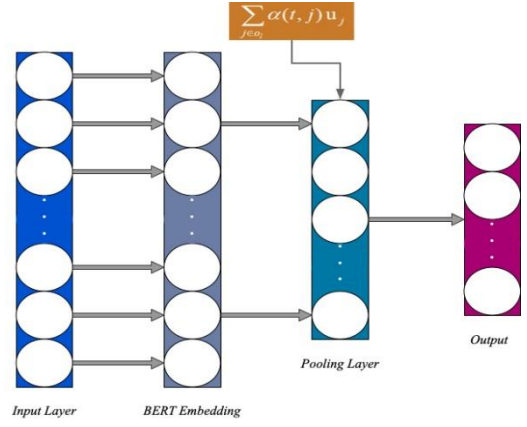

Fig. 4: Illustration of BERT-based attentive aggregation embedding representation learning framework

With the attentive aggregation embedding representation learning, the final group profile contains more abundant feature and enhances the influence of mavens of group. In general group modeling task, it is to obtain an embedding vector for each group to predict its preference on an items. We learn the dynamic aggregation strategy from data in our work and group as embedding, which can be defined as:

$$g_l(t) = f_a(v_t, \{u_j\}_{j \in o_l}) \quad (2)$$

where $g_l(t)$ denotes the embedding of group $g_l$, $o_l$ contains the group member indexes of group $g_l$ and $f_a$ presents the aggregation function.

In our aggregation representation learning framework, we aggregate the generated group vector $g_l'$ by BERT and mavens embedding vector based on attention mechanisms. Our group profile modeling consists of two components, namely attentive mavens embedding and group preference BERT embedding, which can be showed as:

$$g_l(t) = \sum_{j \in o_l} \alpha(t, j) u_j + g_l' \quad (3)$$

### D. General Framework with NCF

NCF is a multi-layer neural network general for item recommendation [37] and has been successfully extended to attribution-based social recommendation [38], review-based product recommendation [39] and group recommendation [11]. These previous NCF work of is as the foundation of our work to some extent. Its main idea is to embedding and item embedding into a dedicated neural network to learn interactive function from data. In our work, we use NCF general framework to learn and predict the interaction between group, user and item. Firstly, given group-item pair $(g_l, v_t)$ or user-item $(u_i, v_t)$, the representation can be obtained by our above representation learning. And then the embeddings vector are fed into a pooling layer and hidden layers and obtain the prediction results at last. Figure 5 illustrates our general NCF framework for solving the group recommendation task. We elaborate each layer and model optimization in our framework below.

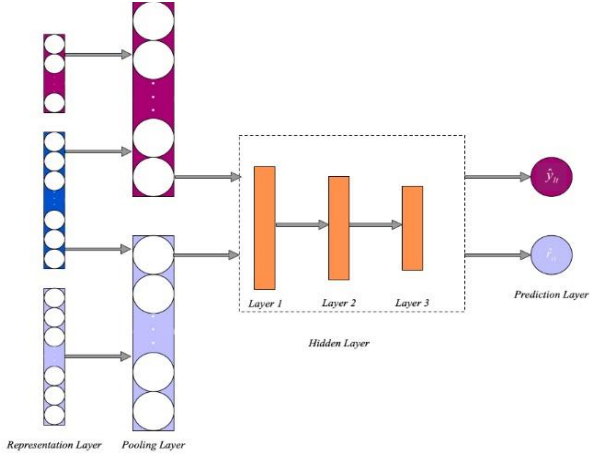

Fig. 5: Illustration of general NCF of interaction learning framework

**Pooling layer:** We assume the input is group-item pair ($g_l$, $v_t$), pooling layer first makes element-wise product on their embeddings and then we concatenate them to the original embedding to form a matrix.

$$e_0 = \varphi_{pooling}(g_l(t), v_t) = \begin{pmatrix} g_l(t) \odot v_t \\ g_l(t) \\ v_t \end{pmatrix} \quad (4)$$

**Hidden layer:** The hidden layer is a stack of fully connected layers and it can capture nonlinear and high order dependencies between users, groups, and items. We use the ReLU function as the activation function.

$$\begin{cases} e_1 = \text{ReLU}(W_1 e_0 + b_1) \\ e_2 = \text{ReLU}(W_2 e_1 + b_2) \\ ... \\ e_n = \text{ReLU}(W_n e_{n-1} + b_n) \end{cases} \quad (5)$$

where $W_n$, $b_n$ and $e_n$ denote the weight matrix, bias vector and output neurons of the n-th hidden layer respectively.

**Prediction layer:** The prediction layer predict the score and the output of the last hidden layer, w represents the weights in prediction layer, on the left side of the formula are the user-item pair ($u_l$, $v_t$) and group-item pair ($g_l$, $v_t$) respectively, which can be showed as:

$$\begin{cases} \hat{r}_{it} = w^T e_n, e_0 = \varphi_{pooling}(u_i, v_t) \\ \hat{y}_{lt} = w^T e_n, e_0 = \varphi_{pooling}(g_l(t), v_t) \end{cases} \quad (6)$$

**Model optimization:** We use the pair learning method to optimize model parameters and opt the regression-based pairwise loss in our group recommendation task, which is a general approach for item recommendation [38]:

$$\gamma_{group} = \sum_{(l,t,s) \in O'} (y_{lts} - \hat{y}_{lts})^2 = \sum_{(l,t,s) \in O'} (y_{lt} - \hat{y}_{ls} - 1)^2 \quad (7)$$

Where $O'$ represents the training set for group recommendation task and ($l$, $t$, $s$) denotes that group $g_l$ has interacted with item $v_t$, but has not interacted with $v_s$ before.

## IV. EXPERIMENTS

In this section, we first introduce the experimental setup, including datasets, evaluation metrics, comparison method. Then we conduct the extensive experiments on two datasets to answer the following questions:

RQ1: How is the effectiveness of our attentive mavens mining method?

RQ2: How does our proposed aggregation representation learning method perform as compared with the two components of embedding?

RQ3: How does our proposed SIAGR method perform as compared with state-of-the-art group recommendation?

### A. Experimental Setup

#### 1) Datasets

We conduct our experiments of two real-world datasets. The second dataset is CAMRa2011 datasets and it is used in [11, 45] and we recommend the movie for households. The final dataset contained 290 groups, 690 users and 7710 items. The average group size is 2.08.

TABLE 1: BASIC STATISTICS OF THE DATASETS

| Dataset | #Groups | #Users | #Items | Avg. Group Size |
|---|---|---|---|---|
| CAMRa2011 | 290 | 690 | 7710 | 2.08 |
| Plancast | 25,447 | 41,065 | 13,514 | 12.01 |

The second dataset is Plancast and it is used in [5, 10]. An event in Plancast consists of a user group and a venue and we regard an event as a group, each user in the event as a group member. In our work, we recommend the venue for the group. The final dataset contained 41,065 users, 25,447 groups and 13,514 items. The average group size is 12.01. Table 1 denotes the basic statistics of the two datasets.

#### 2) Evaluation Metrics

To evaluate the performance of our proposed method for group recommendation, we evaluate the performance of the top-N recommendation. We employed the widely used metric in our experiment and namely Hit Ratio (HR) and Mean Reciprocal Rank (MRR). The two metric are used in [10, 11, 46].

The metric Hits Ratio and Mean Reciprocal Rank are defined as follows:

$$HR@n = \frac{\#hit@n}{|D_{test}|}$$
$$MRR = \frac{1}{|D_{test}|} \sum_{(g,v) \in D_{test}} \frac{1}{rank(v)} \quad (8)$$

where #$hit@n$ represents the number of hits in the test set and $|D_{test}|$ is the total number of test cases in the test set and the value is higher, the performance is well.

#### 3) Comparison Methods

In our experiment, we compared five recommendation methods to justify the effectiveness of our SIAGR method in the whole and the compared methods are AGREE [11], PIT [26], COM [5], NCF+AVG [7] and NCF+LM [8]. Owing to our method is under the framework of NCF, we compare the NCF with static strategies baseline method. AGREE is similar to our method but we employ the more effective aggregation representation learning method for group modeling. PIT is also to learn the personal impact weight for each other and we compare the method to justify the effectiveness of our attentive weights. COM is a probabilistic model and the comparison to show the effectiveness of our deep learning method.

**Attentive group recommendation (AGREE):** AGREE employs attention mechanism to learn the interaction group-item and user-item, but it does not consider the interaction of user-user. Our method is different from this method in representation learning and group modeling and consider fully the interaction of user-user, user-item and group-item.

**Personal impact topic model (PIT):** PIT is an author-topic model and chooses the relatively large influence score as the representation of group and based on the preference to choose the topic, the topic generates the recommendation results.

**Consensus model (COM):** COM is the state-of-the-art group recommendation method and the probabilistic model generates the recommendation activities for group.

**Neural collaborative filtering with averaging strategy (NCF+AVG):** NCF+AVG averages the preference scores of group members as the group preference under the framework of neural collaborative filtering. The method assume the influence of group members are equal.

**Neural collaborative filtering with least-misery strategy (NCF+LM):** NCF+LM uses the least misery strategy and use the minimum score of group members as the group preferences under the framework of neural collaborative filtering. The method assume the least satisfied members have large influence in group decision-making.

### B. Experimental Results

#### 1) Effect of Attention Mechanism (RQ1)

As noted previously, group members have different influence in the group decision-making and mavens of group have a large influence. Figure 6 visualizes the influence weight using PIT and SIAGR in randomly-chosen groups and we can see it clearly that the two methods learn both the influence weight for each group member. In Group A, uer-332 of four users has large weight and it as the maven of group A. However, PIT continues to consider that user-332 has large weight in group B, whereas SIAGR can determine the maven of group effectively for group decision-making. Therefore, our attentive mavens mining method based on attention mechanism can capture the dynamic influence weight for group decision-making.

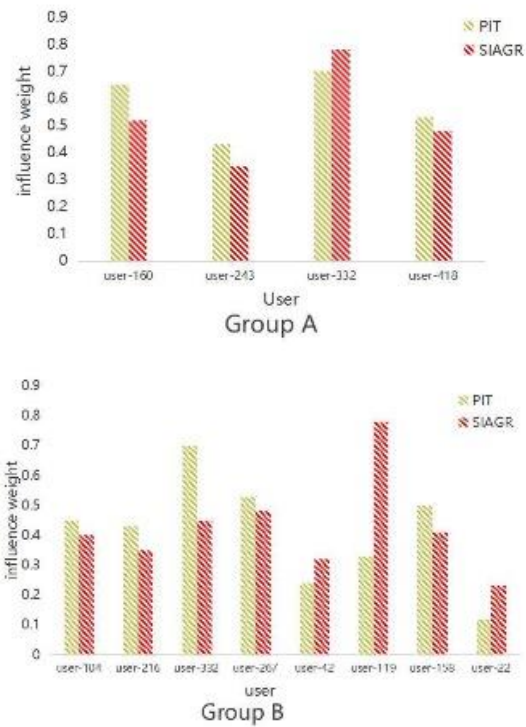

Fig. 6: Visualization for the influence weight by PIT and SIAGR

#### 2) Importance of Aggregation Components (RQ2)

To further understand the importance of aggregation the BERT-based group embedding and attentive mavens embedding, we conduct the experiment. SIAGR-G represents the BERT-based group embedding, SIAGR-M represents the attentive mavens embedding and SIAGR is our attentive aggregation representation learning method. Figure 7 and figure 8 show the results of two simplified components and aggregation components on the two datasets. The explanation of results are as follows: our SIAGR aggregates the SIAGR-G and SIAGR-M as the representation for group modeling. So the performance of our SIAGR is better than the two components.

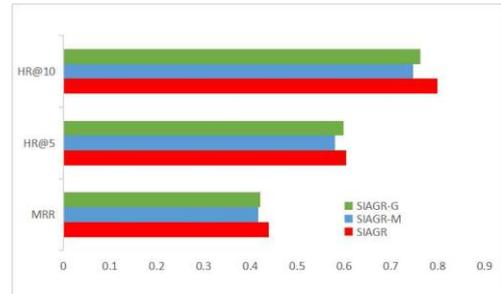

Fig. 7: Importance of aggregation components in dataset 1

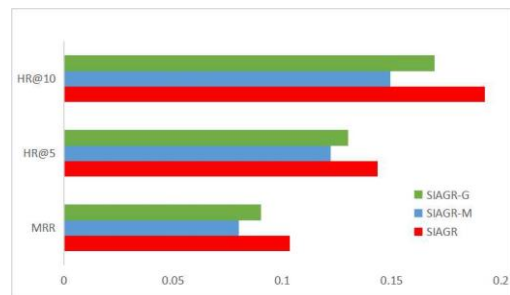

Fig. 8: Importance of aggregation components in dataset 2

#### 3) Overall Performance comparison (RQ3)

We compare the performance of our SIAGR with the relevant baseline and state-of-the-art method for group recommendation task. Figure 9 and figure 10 show the results of different six methods on the two datasets receptively. The relevant observations are as follows: our SIGAR for group recommendation is similar to AGREE but we use the different aggregation representation learning method and our method is better than AGREE in the two datasets. So the effectiveness of our SIAGR in representation learning is justified. Meanwhile, our SIAGR compares the performance with the COM, NCF+AVG and NCF+LM and the results have a better improvement.

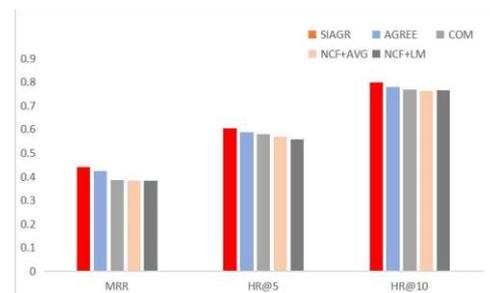

Fig. 9: Overall performance comparison in dataset 1

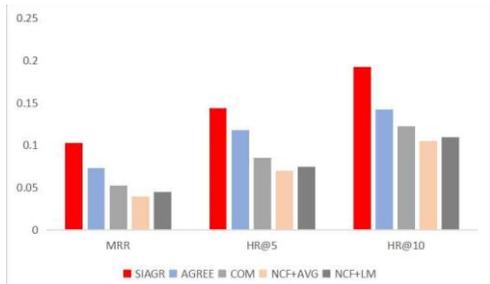
Fig. 10: Overall performance comparison in dataset 2

## V. CONCLUSIONS AND FUTURE WORK

In this work, we focus on the issue of preference aggregation for group recommendation task. We propose a novel attentive aggregation representation learning method based on sociological theory. Firstly, we analyze the influence of group members using social identity theory two-step flow theory. Secondly, based on the sociological theory and we apply attention mechanism to develop an attentive mavens mining method for group. Thirdly, we employ BERT to learn the representation of group and aggregate the group embedding and mavens' preference to perform the final group profile modeling. Lastly, we suggest the recommendation goals under the framework of NCF and our proposed method demonstrates impressive performance in the experimental results. Our work has several limitations for random group and the cold-start problem for group is not fully considered. Some of which can serve as fruitful areas for future areas and we see many possible avenues for future work.